\documentclass[11pt]{article}
\usepackage{amsmath,amssymb}
\usepackage{epsfig,color}
\usepackage{amsmath,amssymb}
\usepackage{amssymb}

\newcommand{\D}{\Delta}

\def\be{\begin{equation}}
\def\ee{\end{equation}}
\def\bea{\begin{eqnarray}}
\def\eea{\end{eqnarray}}
\def\ba{\begin{array}}
\def\ea{\end{array}}
\def\bi{\begin{itemize}}
\def\ei{\end{itemize}}

\renewcommand{\baselinestretch}{1.5}
\def\xxx#1           {{\sf hep-th/#1} }

\def\npb#1(#2)#3     {Nucl. Phys. {\bf B#1} (#2) #3 }
\def\rep#1(#2)#3     {Phys. Rept. {\bf #1} (#2) #3 }
\def\pla#1(#2)#3     {Phys. Lett. {\bf A#1} (#2) #3 }
\def\plb#1(#2)#3     {Phys. Lett. {\bf B#1} (#2) #3 }
\def\prl#1(#2)#3     {Phys. Rev. Lett. {\bf #1} (#2) #3 }
\def\prd#1(#2)#3     {Phys. Rev. {\bf D#1} (#2) #3 }
\def\ap#1(#2)#3      {Ann. Phys. {\bf #1} (#2) #3 }
\def\rmp#1(#2)#3     {Rev. Mod. Phys. {\bf #1} (#2) #3 }
\def\cmp#1(#2)#3     {Comm. Math. Phys. {\bf #1} (#2) #3 }
\def\mpla#1(#2)#3    {Mod. Phys. Lett. {\bf A#1} (#2) #3 }
\def\ijmp#1(#2)#3    {Int. J. Mod. Phys. {\bf A#1} (#2) #3 }
\def\cqg#1(#2)#3     {Class. Quant. Grav. {\bf #1} (#2) #3 }
\def\am#1(#2)#3      {Adv. Math. {\bf #1} (#2) #3 }
\def\im#1(#2)#3      {Invent. Math. {\bf #1} (#2) #3 }
\def\jhep#1(#2)#3    {JHEP {\bf #1} (#2) #3 }
\def\npps#1(#2)#3    {Nucl. Phys. Proc. Suppl. {\bf #1} (#2) #3 }
\def\jgp#1(#2)#3     {J. Geom. Phys. {\bf #1} (#2) #3 }
\def\atmp#1(#2)#3    {Adv. Theor. Math. Phys. {\bf #1} (#2) #3}
\def\lmp#1(#2)#3     {Lett. Math. Phys. {\bf #1} (#2) #3}
\begin{document}
%%%%%%%%%
% Front page here
\thispagestyle{empty}
%\vspace*{1cm}
\null\vskip-24pt \hfill AEI-2001-055 \vskip-10pt \hfill
LAPTH-849/01
%\vskip-10pt \hfill IFUM-FT-683
\vskip-10pt \hfill {\tt hep-th/0105254}
\vskip0.2truecm
\begin{center}
\vskip 0.2truecm {\Large\bf
%\titleline
On Non-renormalization and OPE in Superconformal Field Theories }\\
\vskip 0.5truecm
%\vfill
{\bf G. Arutyunov$^{*,**}$\footnote{email:{\tt agleb@aei-potsdam.mpg.de}}, B.
Eden$^{\ddagger }$\footnote{email:{\tt burkhard@lapp.in2p3.fr}}, %A. C.
%Petkou$^{\dagger}$ \footnote{email:{\tt Anastasios.Petkou@mi.infn.it}},
E. Sokatchev$^{\ddagger}$ \footnote{email:{\tt sokatche@lapp.in2p3.fr} \\
$^{**}$On leave of absence from Steklov Mathematical Institute, Gubkin str.8,
117966, Moscow, Russia }
}\\
\vskip 0.4truecm
%\addresses
$^{*}$ {\it Max-Planck-Institut f\"ur Gravitationsphysik,
Albert-Einstein-Institut, \\
Am M\"uhlenberg 1, D-14476 Golm, Germany}\\
\vskip .2truecm $^{\ddagger}$ {\it Laboratoire d'Annecy-le-Vieux de Physique
Th{\'e}orique\footnote{UMR 5108 associ{\'e}e {\`a} l'Universit{\'e} de Savoie}
LAPTH, Chemin de
Bellevue, B.P. 110, F-74941 Annecy-le-Vieux, France} \\
%\vskip .2truecm $^{\dagger}${\it Dipartimento di Fisica dell'Universita di
%Milano and I.N.F.N. Sezione di Milano,
%\\ via Celoria 16,
%20133 Milano, Italy}\\
\end{center}

\vskip 1truecm \Large
%\noindent
\centerline{\bf Abstract} \normalsize The OPE of two $N=2$ $R$-symmetry current
(short) multiplets is determined by the possible superspace three-point
functions that two such multiplets can form with a third, a priori long
multiplet. We show that the shortness conditions on the former put strong
restrictions on the quantum numbers of the latter. In particular, no anomalous
dimension is allowed unless the third supermultiplet is an $R$-symmetry
singlet. This phenomenon should explain many known non-renormalization
properties of correlation functions, including the one of four stress-tensor
multiplets in $N=4$ SYM$_4$.
\newpage
\setcounter{page}{1}\setcounter{footnote}{0}

\section{Introduction}

Conformal field theories in four dimensions have been the subject
of intensive investigations since the last few years. This surge
of interest has been primarily motivated by the search for
confirmation of the celebrated AdS/CFT correspondence
\cite{M}-\cite{W}. Various methods have been used to study
correlation functions of a special type of CFT operators (called
``short" or ``analytic") in $N=4$ supersymmetric Yang-Mills theory
(SYM), which were then compared to their AdS supergravity
counterparts. These include calculations in supergravity
\cite{FrMaMaRa}-\cite{ArFr3p} and in  perturbative SYM
\cite{HFS}-\cite{PenSaZa}, instanton calculations
\cite{Instanton}, as well as different studies of the general
(non-perturbative) properties of CFT correlators
\cite{hw1}-\cite{FZ}.

Most of the results obtained so far concern the so-called
``non-renormalization" of correlation functions. The earliest
tests for the AdS/CFT correspondence were carried out with the
simplest correlators, namely two- or three-point functions of
short operators. Their form is predicted by superconformal
covariance up to an overall factor (``coupling"). Using various
methods it was shown that these correlators are non-renormalized
and being extracted from supergravity \cite{FrMaMaRa,lmrs,ArFr3p}
match their CFT partners
\cite{HFS,HoSoWe,EHW,BKRS1,Skiba,PenSaZa}. Similar results have
been obtained for special classes of $n$-point correlators called
``extremal" \cite{DHoFrMaMaRa,BiaKov,EdHoScSoWe} and
``next-to-extremal"
\cite{EdHoScSoWe}-\cite{EdHoScSoWeSub}.\footnote{Note that another
special class of so-called ``near-extremal" correlators exhibits
the interesting property of factorization of the amplitude
\cite{HEFP,HokerPioline}.}

More elaborate tests involved four-point functions of short
operators. These correlators are in general non-trivial because
superconformal invariance leaves some functional freedom (see,
e.g., \cite{Pickering} for an abstract study of the correlator of
four stress-tensor multiplets of $N=4$ SYM). Nevertheless, their
functional form can be further restricted by evoking field-theory
dynamical arguments (Intriligator's insertion procedure \cite{I}).
This phenomenon was called ``partial non-renormalization" in
\cite{EPSS} and it was confirmed by comparing to explicit
perturbative \cite{EHSSW}, instanton \cite{BKRS1} and supergravity
\cite{ArutFrol} calculations.

Alternatively, one may explain the special properties of four-point functions
of short operators starting from the operator product expansion (OPE). Indeed,
a four-point function can be viewed as the convolution of two OPEs (a double
OPE). The OPE of two short multiplets may include short as well as long
multiplets (see the classification in \cite{dobPet}). The dimension of the
short multiplets is protected by superconformal invariance whereas that of the
long multiplets may become anomalous due to radiative corrections. This fact
was used in \cite{FZ} to argue that renormalization effects in CFT can be
expected to take place whenever a long multiplet can appear in the OPE of two
short multiplets. The simplest example of this phenomenon involves the Konishi
multiplet \cite{Konishi}. Its occurrence has been demonstrated and its
anomalous dimension has been calculated perturbatively
\cite{AnGrJo,BKRS1,BKRS2}. At the same time, such simple arguments do not
explain the absence of anomalous dimension of certain operators. Among them one
finds \cite{AFP1,AFP2} a scalar double trace operator $O_{20}$ arising in the
OPE of the lowest dimensional scalars from the stress-tensor (supercurrent)
multiplet in $N=4$ SYM. Recently it was shown \cite{AEPS} that the
non-renormalization of $O_{20}$ follows from the partial non-renormalization of
\cite{EPSS}. {}From the free theory one finds several double trace operators in
the {\bf 20} of $SU(4)$ which are in general expected to mix under the
renormalization group flow. The mixing problem was addressed in
\cite{BKRSkonishi}, where the perturbative properties of the Konishi multiplet
were analyzed and the same conclusion about the $O_{20}$ was again reached at
the perturbative level.

The OPE techniques used so far have been based on ordinary space-time. In a
supersymmetric theory it is natural to study OPEs in superspace, in order to
benefit from the manifest supersymmetry \cite{HWOPE}. In Section 2 of this
letter we discuss the OPE of two short (Grassmann and harmonic analytic
\cite{GIO}-\cite{HowHar},\cite{hw1}) multiplets. Its content is determined by
the possible three-point functions having two short supermultiplet legs and one
general leg. For simplicity we restrict ourselves to the case where the short
multiplets are $N=2$ $R$-symmetry current multiplets, but our argument can
easily be adapted to other types of $N=2$ or $N=4$ multiplets. We explain how
such three-point functions can be constructed in $N=2$ harmonic superspace
\cite{hh,TheBook}.\footnote{Many papers have been devoted to the construction
of superconformal two- and three-point functions in superspace. These include
the case of $N=1$ chiral superfields \cite{ConWest,DolOsb} as well as a method
valid for all types of $N=1$ superfields \cite{Park, Osborn}. A construction of
analytic $N=2$ and $N=4$ two- and three-point functions is presented in
\cite{hw1} (see also \cite{HoSoWe}). Ref. \cite{KuzTheis} discusses the $N=2$
case in the context of three protected (current or stress-tensor) operators.
However, no explicit results are available for $N>1$ in the more general case
where one of the operators in the three-point function is generic.} We show
that the combined requirements of shortness (Grassmann analyticity and $R$
symmetry irreducibility, or H-analyticity) at two ends lead to strong
restrictions on the supermultiplet at the third, general end. We find that the
supermultiplets which are allowed to have anomalous dimension must be singlets
under the $R$-symmetry group. This provides a simple explanation of the
surprising absence of anomalous dimension of operators like the aforementioned
$O_{20}$.

Extrapolating the above $N=2$ considerations to the OPE of two primary
operators from the $N=4$ stress-tensor multiplet, we expect that this OPE can
contain superconformal primary operators with a non-vanishing anomalous
dimension {\it only} if it is a singlet of $SU(4)$. In Section 3 we demonstrate
that this conclusion is compatible with the known structure of the four-point
correlation functions of this type.

The results presented here are still preliminary, in the sense that we only
find necessary conditions for the existence of the three-point functions
considered. A systematic study of this class of three-point functions is under
way and will be presented in a forthcoming publication.

\section{Three-point functions involving two short $N=2$ operators}

The simplest example of a short gauge invariant composite operator
in $N=2$ supersymmetry is a bilinear combination of $N=2$ matter
(hypermultiplet) superfields:
\begin{equation}\label{1.1}
  L^{++}(x,\theta^+,\bar\theta^+,u) = {\rm
  Tr}[q^+(x,\theta^+,\bar\theta^+,u)]^2\;.
\end{equation}
Off shell it satisfies the so-called Grassmann (G-)analyticity
condition \cite{GIO}-\cite{TheBook}. It means that the superfields
in (\ref{1.1}) depend only on one half of the Grassmann variables
obtained by projecting the $SU(2)$ ($R$ symmetry) index by
harmonic variables:
\begin{equation}\label{1.2}
 \theta^{+\alpha} = u^+_i\theta^{i\alpha},\quad
\bar\theta^{+\dot\alpha} = u^+_i\bar\theta^{i\dot\alpha}\;.
\end{equation}
The charges in (\ref{1.1}), (\ref{1.2}) correspond to the $U(1)$
subgroup of $SU(2)$.

On shell the superfield $L^{++}$ satisfies an additional condition
restricting its harmonic dependence,
\begin{equation}\label{1.3}
  D^{++}L^{++}=0\;.
\end{equation}
Here the harmonic derivative
\begin{equation}\label{1.4}
  D^{++}= u^{+i} \frac{\partial}{\partial u^{-i}} - 2i \theta^{+}\sigma^\mu \bar
\theta^{+} \frac{\partial}{\partial x^\mu}
\end{equation}
plays the r\^ole of the raising operator of $SU(2)$. In this sense
eq. (\ref{1.3}) can be regarded as the condition for $SU(2)$
irreducibility defining $L^{++}$ as the highest weight state of a
triplet irrep of $SU(2)$. An alternative interpretation is
obtained by parametrizing the harmonic coset $S^2\sim SU(2)/U(1)$
by a complex variable and treating  eq. (\ref{1.3}) as the
condition for Cauchy-Riemann analyticity (H-analyticity
\cite{HowHar}). The solution of the combined G- and H-analyticity
constraints is a {\it short superfield} (known is the $N=2$ linear
or tensor multiplet \cite{aD111}) whose bosonic content is as
follows:
\begin{eqnarray}
  L^{++}&=& L^{ij}(x)u^+_iu^+_j + (\theta^+)^2 M(x) + (\bar\theta^+)^2 \bar M(x)\nonumber\\
  && + 2i\theta^+\sigma^\mu\bar\theta^+ [J_\mu(x) + \partial_\mu
L^{ij}(x)u^+_iu^-_j ] + (\theta^+)^4\square
L^{ij}(x)u^-_iu^-_j\,.\label{1.5}
\end{eqnarray}
Note that the vector in (\ref{1.5}) has a triplet part
$\partial_\mu L^{ij}(x)$ expressed in terms of the scalar, and a
singlet part $J_\mu(x)$ which must be conserved,
\begin{equation}\label{1.6}
  \partial^\mu J_\mu(x) =0\;.
\end{equation}
Thus, the linear multiplet is also the  $R$-symmetry  current multiplet in an
$N=2$ superconformal theory. Superconformal invariance and the above
constraints fix the dimension and $R$ weight of the supermultiplet (this is
typical for short superconformal multiplets, see \cite{dobPet}). Denoting by
\begin{equation}\label{1.6''}
  {\cal D}=(d,s,r,a)
\end{equation}
the superconformal quantum labels dimension $d$, spin $s$, $R$
weight $r$ and $SU(2)$ Dynkin label $a$ ($a$ equals the harmonic
$U(1)$ charge), we find
\begin{equation}\label{1.6'}
  {\cal D}_{L^{++}} = (2,0,0,2)\;.
\end{equation}

Now, in the theory of $N=2$ hypermultiplets interacting with $N=2$
SYM one can consider the four-point correlation function
\begin{equation}\label{1.7}
\langle L^{++}(1)L^{++}(2)L^{++}(3)L^{++}(4)\rangle\;.
\end{equation}
A certain class of such theories (e.g., $N=4$ SYM written in terms
of $N=2$ superfields) are known to be finite and hence
superconformal. Then one can apply the standard approach of
(super)conformal OPE and write down the correlator (\ref{1.7}) in
the form of a double OPE (see, e.g., \cite{FGG,DPPT,Pisa,FP}):
\begin{eqnarray}\label{1.8}
&& \langle L^{++}(1)L^{++}(2)L^{++}(3)L^{++}(4)\rangle = \\
\nonumber
&&  \sum_{\cal D} \int_{5,5'}\langle L^{++}(1)L^{++}(2)
   {\cal O}_{ {\cal D}}(5)\rangle\ \langle{\cal O}_{\cal D}(5)
   {\cal O}_{\cal D}(5')\rangle^{-1} \ \langle
   {\cal O}_{ {\cal D}}(5') L^{++}(3)L^{++}(4)\rangle\;. \nonumber
   \end{eqnarray}
Here ${\cal O}_{\cal D}$ denotes an operator in a superconformal
irrep with labels (\ref{1.6'}). The sum in (\ref{1.8}) goes over
all possible such irreps.

It is then clear that the spectrum of this OPE as well as the
content of the double OPE (\ref{1.8}) is determined by the
possible superconformal three-point functions. When we say
``possible" we have in mind that at points 1 and 2 such a function
must satisfy the G- and H-analyticity conditions above. As we
shall see shortly, the combination of these conditions with
superconformal covariance imposes strong restrictions on the
allowed irreps ${\cal O}_{ {\cal D}}(3)$ at point 3 and hence on
the content of the OPE (\ref{1.8}).

So, our task will be to study the three-point functions
\begin{equation}\label{1.10}
  G_{ {\cal D}} = \langle L^{++}(1)L^{++}(2)
   {\cal O}_{ {\cal D}}(3)\rangle
\end{equation}
and to find out which irreps $ {\cal D}$ can appear at point 3. To
start with, we remark that because of G-analyticity at points 1
and 2 the function $G_{ {\cal D}}$ depends on two ``half spinor"
Grassmann coordinates $\theta^{+\alpha,\dot\alpha}_{1,2} =
u^+_{1,2\; i}\theta^{i\;\alpha,\dot\alpha}_{1,2}$ and on one
``full spinor" $\theta^{i\;\alpha,\dot\alpha}_3$. At the same
time, the superconformal algebra contains two ``full spinor"
generators, that of Poincar\'e supersymmetry
$Q^{i\;\alpha,\dot\alpha}$ and that of conformal supersymmetry
$S^{i\;\alpha,\dot\alpha}$. These generators act on the Grassmann
coordinates as shifts (non-linear in the case of $S$
supersymmetry), so the three-point function (\ref{1.10})
effectively depends on $1/2+1/2+1- (1+1) = 0$ invariant
combinations of Grassmann variables. In other words, given the
lowest component (obtained by setting
$\theta^+_1=\theta^+_2=\theta^i_3=0$) of the three-point function
(\ref{1.10}), $Q$ and $S$ supersymmetry allow one to find its {\sl
unique completion} to a full superfunction. Another implication of
this fact is the impossibility to have $R$ weight at point 3.
Indeed, the only superspace coordinates carrying $R$ weight are
the Grassmann ones, and the above counting shows that there is no
invariant combination of the available $\theta$'s.

Let us first examine the lowest component $G_{ {\cal D}}(\theta=0)$. Since the
operators at points 1 and 2 are $SU(2)$ triplets, the allowed $SU(2)$ irreps at
point 3 are $\underline 3 \times \underline 3\rightarrow \underline 1 +
\underline 3 + \underline 5$. Using the notation $[2,2,a]$ for an $SU(2)$
structure with charge ($SU(2)$ Dynkin label) 2 at points 1,2 and charge $a=0,2$
or 4 at point 3, we can write:
\begin{equation}\label{1.11}
  G_{ {\cal D}}(\theta=0) = [2, 2, a] \frac{V^{\{\mu_1}\ldots V^{\mu_s\}}}{(x_{12}^2)^{2-\frac{  d-s}{2}}
  (x_{13}^2x_{23}^2)^\frac{  d-s}{2}}\;.
\end{equation}
Here
$$
V^\mu = \frac{x_{13}^\mu}{x_{13}^2} - \frac{x_{23}^\mu}{x_{23}^2}
$$
is a vector with a homogeneous conformal transformation law (see,
e.g., \cite{OP}) and $\{\}$ means a symmetrized traceless
product.\footnote{It is well-known that only symmetric traceless
tensors can appear in the OPE of two scalar operators \cite{FGG}.}
The symbol $[2, 2, a]$ denotes the isospin index structure
associated to each allowed value of $a$. Since the $SU(2)$ indices
at points $1,2$ are projected by harmonics, we find it convenient
to introduce an auxiliary harmonic variable at point 3 as well.
Using the short-hand notation $(12)\equiv u^{+i}_1u^+_{2i}$ for
$SU(2)$ invariant harmonic contractions, we can write down the
three possible isospin structures in (\ref{1.11}) as follows:
\begin{equation}\label{1.12}
  [2,2,0]=(12)^2\;; \qquad [2,2,2]= (12)(13)(23)\;; \qquad [2,2,4]=(13)^2
  (23)^2\;.
\end{equation}

Next we start the completion of the lowest component (\ref{1.11})
to a full superfunction. As explained above, $Q$ and $S$
supersymmetry allow one to do this in a unique way. We find it
convenient to first exploit translation invariance and $Q$
supersymmetry and fix a frame in which $x_3=\theta_3=0$. Such a
choice is stable under (super)conformal transformations, so we can
proceed by just making an $S$ supersymmetry transformation at
points 1 and 2. In this letter we shall restrict ourselves to the
first non-trivial term in the $\theta$ expansion of $G_{ {\cal
D}}$. $R$ weight preservation tells us that these terms must be of
the type $\theta\bar\theta$. So, we write schematically
\begin{eqnarray}
  G_{ {\cal D}}&=& [2, 2, a] \frac{V^{\{\mu_1}\ldots V^{\mu_s\}}}{(x_{12}^2)^{2-\frac{  d-s}{2}}
  (x_{1}^2x_{2}^2)^\frac{  d-s}{2}} \label{1.13}\\
  && + 2i \left\{\theta^+_1 \sigma_\nu\bar\theta^+_1
  \rho^{[0,2,a]\; \nu\{\mu_1\ldots\mu_s\}}_{11}
  +\theta^+_1 \sigma_\nu\bar\theta^+_2 \rho^{[1,1,a]\;\nu\{\mu_1\ldots\mu_s\}}_{12} + (1\leftrightarrow 2)\right\}
  + O(\theta^4)\nonumber
\end{eqnarray}
In order to determine the vector/tensors
$\rho^{\nu\{\mu_1\ldots\mu_s\}}$ we need to make a linearized $S$
supersymmetry transformation at points 1 and 2. The G-analytic
coordinates transform as follow \cite{aG9,TheBook} (only the
right-handed parameter $\bar\eta_{i\dot\alpha}$ is shown):
\begin{eqnarray}
  \delta_S x^{\dot\alpha\alpha}&=& -4i(x^{\alpha \dot\beta}
\bar\eta^i_{\dot\beta} u^-_i) \bar\theta^{+ \dot\alpha} \nonumber\\
  \delta_S \theta^{+\alpha}&=& x^{\alpha \dot\beta} \bar\eta^i_{\dot\beta}
u^+_i \nonumber\\
  \delta_S \bar\theta^{+\dot\alpha}&=&  O(\bar\theta^2) \nonumber\\
  \delta_S u^+_i &=& 4i(u^+_j \bar\eta^j_{\dot\beta} \bar\theta^{+ \dot\beta})
u^-_i \;, \qquad \delta_S u^-_i = 0\;.    \label{1.14}
\end{eqnarray}
Further, the linear multiplet transforms with a fixed weight
\cite{TheBook},
\begin{equation}\label{1.15}
  \delta_S G_{ {\cal D}} = -8i \bar\eta_{\dot\beta i}(\bar\theta_1^{+\dot\beta}
u^{-i}_1 + \bar\theta_2^{+\dot\beta} u^{-i}_2) \, G_{ {\cal D}}
\end{equation}
(recall that we have set $x_3=\theta_3=0$, so we need not consider the
superconformal transformation at point 3). {}From (\ref{1.13})--(\ref{1.15})
one finds linear equations for the $\rho$'s. The solution for $\rho_{11}$ is
\begin{eqnarray}\label{1.16}
  && \rho^{[0,2,a] \, \nu\{\mu_1\ldots\mu_s\}}_{11} =
  \frac{2}{(x_{12}^2)^{2-\frac{  d-s}{2}} (x_{1}^2 x_{2}^2)^\frac{  d-s}{2}}
  \biggl[ \frac{ [1,3,a]}{(12)}
  \, \bigl( -2 \, \frac{x_1^\nu}{x_1^2} \bigr) \, V^{\{\mu_1}\ldots
V^{\mu_s\}} + \\
  && + \frac{(1^-2) \, [2,2,a]}{(12)}
  \left\{ \bigl( \bigl( \frac{  d-s}{2}-2 \bigr)
  \frac{x_{12}^\nu}{x_{12}^2} - \frac{  d-s}{2} \,
  \frac{x_1^\nu}{x_{1}^2} \bigr)
  \, V^{\{\mu_1}\ldots V^{\mu_s\}} + \frac{s}{2 x_1^2}
  \, I^{\nu\{\mu_1}(x_1) \, V^{\mu_2} \ldots V^{\mu_s\}} \right\} \biggr] \nonumber
\end{eqnarray}
and similarly for the remaining ones. Here $I^{\nu\mu}(x) = \eta^{\nu\mu} - 2
{x^\nu x^\mu}/{x^2}$ is the inversion tensor;  $(1^-2)\equiv u^{-i}_1 u^+_{2i}$
and
\begin{equation}\label{1.161}
   [1,3,0]=0\;; \qquad [1,3,2]= \frac{1}{2}(12)(23)^2\;;
   \qquad [1,3,4]=(13)(23)^3\;.
\end{equation}

So far we have not obtained any restrictions on the allowed values
of the dimension and spin at point 3. However, there is one
constraint that we have not yet taken into account, namely,
H-analyticity (or $SU(2)$ irreducibility) (\ref{1.3}) at points 1
and 2. It manifests itself in two ways. Firstly, no harmonic
singularities of the type $1/(12)$ are allowed. This is the
condition of analyticity on the harmonic sphere $S^2\sim
SU(2)/U(1)$ which is equivalent to having short (polynomial)
harmonic dependence and hence $SU(2)$ irreducibility. Secondly,
the vector/tensor (\ref{1.16}) plays the r\^ole of the vector
component of a linear multiplet (\ref{1.5}) considered as a
function of $x_1,\theta^+_1$ and $u^\pm_1$. As such, it may have a
triplet part (with respect to the harmonics at point 1) and a
singlet part. According to (\ref{1.6}), the latter must be
conserved.

The implementation of these two constraints depends on the value
of the charge $a$ at point 3.

(i) $a=0$ (singlet operator at point 3). {}From (\ref{1.12}), (\ref{1.16}) and
(\ref{1.161}) we see that there appear no harmonic singularities. Further, in
this case (\ref{1.16}) can be rewritten as follows:
\begin{equation}\label{1.17}
  \rho^{[0,2,0] \, \nu\{\mu_1\ldots\mu_s\}}_{11} =
  (1^-2)(12) \, \partial_1^\nu \left\{ \frac{1}{(x_{12}^2)^{2-\frac{  d-s}{2}} (x_{1}^2 x_{2}^2)^\frac{  d-s}{2}} \, V^{\{\mu_1}\ldots V^{\mu_s\}} \right\}
\end{equation}
The harmonic factor $(1^-2)(12) = u^{-i}_{1}u^{+j}_1 u^+_{2i}u^+_{2j}$
corresponds to an $SU(2)$ triplet (the harmonics commute). Thus, the triplet
vector/tensor in (\ref{1.17})  has the form required by (\ref{1.5}). The
absence of a singlet vector/tensor means that the constraint (\ref{1.6}) does
not apply here. The conclusion is that H-analyticity does not imply any new
restrictions on the allowed irrep ${ {\cal D} }$ at point 3.\footnote{At least,
this is it what we can say by terminating the $\theta$ expansion at the level
shown in (\ref{1.13}). Whether the higher levels bring in some new constraints
or not can only be decided after constructing the complete three-point
superfunction. This will be done elsewhere.} The only constraint originates
from the unitarity bound
\begin{equation}\label{UB}
  d \geq s+2\ \ \mbox{for}\ \ s\geq 1\quad \mbox{or}\quad d\geq 1\ \
  \mbox{for}\ \ s=0\;.
\end{equation}

We should stress that the superconformal irreps of the type ${\cal
D}=(d,s,0,0)$ allowed in this case are in general {\it long
multiplets}.

  (ii) $a=2$ (triplet operator at point 3). This time we find
  \begin{eqnarray}
&& \rho^{[0,2,2] \, \nu\{\mu_1\ldots\mu_s\}}_{11} = \frac{1}{2}
[(1^-2)(13)+ (12)(1^-3)] (23) \, \partial_1^\nu \, \left\{
\frac{1}{(x_{12}^2)^{2-\frac{  d-s}{2}} (x_{1}^2 x_{2}^2)^\frac{
d-s}{2}}
\, V^{\{\mu_1}\ldots V^{\mu_s\}} \right\} \nonumber\\
&& - \frac{1}{2} \, \frac{(23)^2}{(x_{12}^2)^{2-\frac{ d-s}{2}}
(x_{1}^2 x_{2}^2)^\frac{  d-s}{2}} \left\{ (  d-s-4) \bigl(
\frac{x_{12}^\nu}{x_{12}^2} - \frac{x_1^\nu}{x_{1}^2} \bigr) \,
V^{\{\mu_1}\ldots V^{\mu_s\}} + \frac{s}{x_1^2}
I^{\nu\{\mu_1}(x_1) V^{\mu_2} \ldots V^{\mu_s\}} \right\}
\nonumber\\ \label{1.18}
\end{eqnarray}
Once again, there are no harmonic singularities and the triplet part of the
vector/tensor (the first line in (\ref{1.18})) is of the form required by
(\ref{1.5}). However, now the vector/tensor has a singlet part (the second,
$u^\pm_1$-independent line in (\ref{1.18})) which is subject to the
conservation condition (\ref{1.6}). This is possible only if
\begin{equation}\label{1.111}
  d=2-s
\end{equation}
 or
\begin{equation}\label{1.18'}
    d=s+4\;.
\end{equation}
The solution (\ref{1.111}) is incompatible with the unitarity
bound (\ref{UB}) unless
\begin{equation}\label{1.181}
  d=2, \qquad s=0\;.
\end{equation}
In fact, this corresponds to having another linear (i.e., short) multiplet at
point 3. The second solution (\ref{1.18'}) is always allowed.

We conclude that the triplet operator at point 3 cannot have an anomalous
dimension. In the special case (\ref{1.181}) this is explained by the fact that
the operator is short, but in the more general case (\ref{1.18'}) this is a new
result.

  (iii) $a=4$ (5-plet operator at point 3). We find:
  \begin{eqnarray}
&& \rho^{[0,2,4] \, \nu\{\mu_1\ldots\mu_s\}}_{11} =
(1^-3)(13)(23)^2 \, \partial_1^\nu \, \left\{
\frac{1}{(x_{12}^2)^{2-\frac{  d-s}{2}} (x_{1}^2 x_{2}^2)^\frac{
d-s}{2}} \,
V^{\{\mu_1}\ldots V^{\mu_s\}} \right\}\nonumber\\
 &&- \frac{(13)(23)^2}{(12)} \,
\frac{1}{(x_{12}^2)^{2-\frac{  d-s}{2}} (x_{1}^2 x_{2}^2)^\frac{
d-s}{2}} \left\{ (  d-s-4) \bigl( \frac{x_{12}^\nu}{x_{12}^2} -
\frac{x_1^\nu}{x_{1}^2} \bigr) \, V^{\{\mu_1}\ldots V^{\mu_s\}}\right. \nonumber\\
&&\phantom{- \frac{(13)(23)^2}{(12)} \,
\frac{1}{(x_{12}^2)^{2-\frac{  d-s}{2}} (x_{1}^2 x_{2}^2)^\frac{
d-s}{2}} }\left. \; + \frac{s}{x_1^2} \, I^{\nu\{\mu_1}(x_1)
V^{\mu_2} \ldots V^{\mu_s\}}\;. \right\}\label{1.19}
\end{eqnarray}
This time we encounter a harmonic singularity. The two terms in
its coefficient are linearly independent, so the only way to
remove the harmonic singularity is to set
\begin{equation}\label{1.000}
  d=4, \quad s=0\;.
\end{equation}

Again, we conclude that the 5-plet operator is severely restricted. Moreover,
the lowest term (\ref{1.11}) becomes singular for the values (\ref{1.000})
since $(x^2)^{-2} \sim \delta(x)$ in Euclidean space \cite{aG24}.

\section{Four-point functions of the N=4 supercurrent multiplet}

Consider now the $N=4$ case. Here the analogue of the $N=2$ linear multiplet
$L^{++}$ is the supercurrent (stress-tensor) multiplet whose component content
can be deduced from  $L=trW^{\{i}W^{j\}}$, where $W^i$ is the $N=4$ on-shell
superfield carrying the irrep $\bf 6$ of the $R$-symmetry group $SU(4)$. A
superconformal primary operator generating $L$ is a scalar $O^I$ of dimension 2
transforming in the irrep $\bf 20$ of $SU(4)$, $I=1,\ldots, 20$.

Our considerations above exploiting the G- and H-analyticity of the three-point
functions at points $1$ and $2$ show in particular that the OPE of two primary
operators from the $N=4$ supercurrent multiplet can contain superconformal
primary operators with a non-vanishing anomalous dimension {\it only} in the
singlet of $SU(4)$. Now we shall demonstrate that this conclusion is compatible
with the known structure of the four-point correlation functions of the fields
$O^I$.

According to \cite{EPSS} the ``quantum'' part of this four-point function
comprising all possible quantum corrections to the free-field result is given
by a {\it single} function $F(v,Y)$ of conformal cross-ratios, which we
conveniently choose to be  $v=\frac{x_{12}^2x_{34}^2}{x_{14}^2x_{23}^2}$ and
$Y=\frac{x_{13}^2x_{24}^2}{x_{14}^2x_{23}^2}$. Under $SU(4)$ the product of two
$O^I$ decomposes as
\begin{equation}\label{irreps}
  {\bf 20}\times {\bf 20}={\bf 1}+{\bf 20}+{\bf 105}+{\bf 84}+{\bf 15}+{\bf 175} \, .
\end{equation}
The ``quantum'' part of the four-point function of the operators $O^I$
projected on the different irreps in (\ref{irreps}) is \bea \label{proj}
\langle O(x_1)O(x_2)O(x_3)O(x_4)\rangle _i
=\frac{1}{x_{12}^4x_{34}^4}P_{i}(v,Y) \frac{vF(v,Y)}{(1-Y)^2} \, . \eea Here
the polynomials $P_{i}(v,Y)$ are \cite{AFP1, AFP2, AEPS}
\begin{align}
\nonumber
&P_{\bf 1}=1-Y-\frac{4}{15}v+\frac{1}{6}Y^2+\frac{2}{15}vY+\frac{1}{60}v^2;&
&P_{\bf 105}=v^2;  \\
\nonumber
&P_{\bf 20}=-\frac{5}{3}v+\frac{5}{3}Y^2+\frac{5}{6}vY+\frac{1}{6}v^2;&
&P_{\bf 15}=4Y-2Y^2-vY; \\
\nonumber
&P_{\bf 84}=3v-\frac{3}{2}vY-\frac{1}{2}v^2;& &P_{\bf 175}=vY\, .
\end{align}

Every irrep $i$ of $SU(4)$ in the OPE of two $O^I$ represents a contribution
from an infinite tower of operators $O_{\D,l}^{i}$, where $\D$ is the conformal
dimension of the operator, $l$ is its Lorentz spin. The corresponding
contribution to the four-point function can then be represented as an expansion
of the type
\begin{equation}
\label{cpwae}
\langle O(x_1)
O(x_2)O(x_3)O(x_4)\rangle_{i}
=\sum_{\Delta, l} a_{\Delta, l}^{i}{\cal H}_{\Delta, l}(x_{1,2,3,4})\; .
\end{equation}
Here ${\cal H}_{\Delta, l}(x_{1,2,3,4})$ denotes the (canonically normalized)
Conformal Partial Wave Amplitude (CPWA) for the exchange of an operator
$O_{\D,l}^{i}$ and $a_{\Delta, l}^{i}$ is a normalization constant. We treat
the CPWA as a double series of the type \cite{HPR,AEPS}
\begin{equation}
\label{gcpwa} {\cal H}_{\Delta, l} = \frac{1}{x_{12}^4 \, x_{34}^4} \,
v^{\frac{h}{2}} \, \sum_{n,m=0}^\infty c^{\Delta, l}_{nm} v^n Y^m \, ,
\end{equation}
where the dimension $\D$ was split into a canonical part $\D_0$ and an
anomalous part $h$: $\D=\D_0+h$.

Let us assign the grading parameter $T=2n+m$ to the monomial $v^nY^m$. As was
shown in \cite{AEPS} the monomials in (\ref{gcpwa}) with the lowest value of
$T$ have $T=\D_0$, where $\D_0$ is the canonical (free-field) dimension of the
corresponding operator.

Comparing (\ref{proj}) and (\ref{cpwae}) one finds, within every fractional
power $v^{\frac{h}{2}}$, the following compatibility conditions \bea
\label{comp} P_{i}\sum_{\Delta, l} a_{\Delta, l}^{j}{\cal
H}_{\Delta,l}(x_{1,2,3,4}) =P_{j}\sum_{\Delta, l} a_{\Delta, l}^{i}{\cal
H}_{\Delta,l}(x_{1,2,3,4}) \, \eea which hold for all pairs $(i,j)$ of irreps
in (\ref{irreps}). Here the sums are taken over operators which have the same
$h$. Thus, eqs. (\ref{comp}) imply non-trivial relations between the CPWAs of
primary operators belonging to the same supersymmetry multiplet(s) with
anomalous dimension $h$.\footnote{The theory we consider does not have any
additional symmetry except the superconformal one and therefore it is natural
to expect that there do not exist two supersymmetry multiplets built upon
different superconformal primaries but with the same anomalous dimension
considered as a function of the coupling.} Only one of these primary operators
is the superconformal primary operator, i.e., it generates under supersymmetry
the whole multiplet, while the others are its descendents.

We can now argue that a superconformal primary operator appears only in the
singlet of $SU(4)$. Indeed, let us choose in (\ref{comp}) the irrep $j$ to be
the singlet. The polynomial $P_{\bf 1}$  is distinguished from the other
$P_i$'s by the presence of a constant term. Suppose that a superconformal
primary operator with a canonical dimension $\D_0$ contributes to the OPE and
transforms in some irrep $i$ which is not a singlet. Due to the constant in
$P_{\bf 1}$, the lowest-order monomials on the r.h.s. of (\ref{comp}) would
have $T=2n+m=\D_0$. Clearly, all the other $P_i$'s always raise the $T$-grading
by at least unity. The lowest dimension operator with canonical dimension
$\D_0'$ in the singlet would have the lowest terms with at least $T=\D_0-1$ (or
lower) to saturate (\ref{comp}). Hence, $\D_0'$ is always lower then $\D_0$,
and therefore the corresponding operator cannot be a supersymmetry descendent
of an operator in the irrep $i$. This shows that anomalous superconformal
primary operators occur only in the singlet.

The relations (\ref{comp}) may be solved for the normalization constants
$a_{\D,l}^{i}$ in terms of the normalization constant of the superconformal
primary operator $a_{\D_{sp},l_{sp}}^{{\bf 1}}$: \bea \label{rel}
a_{\D,l}^{i}=\lambda_{\D,l}^{i}\cdot a_{\D_{sp},l_{sp}}^{{\bf 1}} \, . \eea On
the other hand, the coefficients $a_{\D,l}^{i}$ are expressed via the
normalization constants of the three-point functions involving two CPOs and
$O_{\D,l}^i$ (we assume that $O_{\D,l}^i$ are canonically normalized). Thus,
equations (\ref{rel}) and (\ref{comp}) are in fact implied by the existence of
the unique superspace structure $\langle L(1)L(2){\cal O}\rangle $, where $\cal
O$ is a (long) superfield whose lowest component is  $O_{\D,l}^i$.

{}From this argument we again deduce that superconformal primaries may belong
to other irreps of $SU(4)$ only if they have a vanishing anomalous dimension.
In the case $h=0$ the Born (free-field) part of the four-point function should
be taken into account which allows the existence of the superconformal
primaries in the other irreps of $SU(4)$. A typical example is the operator
$O^I$. Another interesting example is provided by the double-trace operator of
dimension 4 in {\bf 20} which appears in the OPE of two $O^I$ \cite{AFP1,AFP2}.

\section{Conclusions}

{\it Summary of the results}

The OPE of two $N=2$ current multiplets is determined by the
possible three-point functions that two such multiplets can form
with another, a priori long multiplet. The shortness conditions on
the former restrict the latter to one of the following
superconformal unitary irreps:
\begin{equation}\label{1.20}
  \begin{array}{llll}
    \mbox{$R$ symmetry singlet (a=0):} & d\geq s+2, & s\geq 1, &r=0\\
                                       & d\geq 1, &s=0, &r=0\\
    \mbox{$R$ symmetry triplet (a=2):} & d=2, & s=0, &r=0 \quad
    \mbox{(current m-t)}\\
                                       &d=s+4, & s\geq 0, &r=0\\
    \mbox{$R$ symmetry 5-plet (a=4):} & d=4, & s=0, &r=0
  \end{array}
\end{equation}
The only case in which an operator with anomalous dimension is allowed in the
OPE is that of the singlet. The dimension of the triplet and the 5-plet is
protected. This new phenomenon is at the origin of many known
non-renormalization theorems.

These necessary conditions for the existence of the three-point function are
derived from the first non-trivial term in the $\theta$ expansion (\ref{1.13}).
In a future publication we shall complete the result by examining the full
superfunction.

{\it Possible further developments}

The generalization of the above simplest example to other $N=2$ and $N=4$
analytic (short) multiplets at points 1 and 2 is straightforward and will be
given elsewhere.

Another interesting question is what happens if one applies Intriligator's
insertion procedure \cite{I} to the double OPE (\ref{1.8}). Presumably, this
should lead to an additional selection rule for the operators in the OPE. For
instance, it is easy to show that analytic multiplets cannot appear in it. This
should explain the difference between the general superconformal result of Ref.
\cite{Pickering} for the four-point function of $N=4$ SYM stress-tensor
multiplets and the dynamical result of Ref. \cite{EPSS}.

We believe that a similar OPE argument may be applied to the extremal  and
next-to-extremal $n$-point correlators, providing an easy way to show their
non-renormalization. It is also tempting to look for an explanation of the
near-extremal factorization phenomenon along the same lines.

Recently non-renormalization properties of certain  multitrace
operators were discussed \cite{Lemes, Maggiore} by using the
twisted formulation of $N=2$ theories. It would be interesting to
understand if this approach is related to ours.

\bigskip\bigskip

{\bf Note added.} Before submitting this publication to the e-archive we saw
the new paper \cite{HWnew} in which it is argued that long multiplets may
appear in the OPE of analytic superfields (cf. Ref. \cite{HWOPE}).

\section*{Acknowledgements}
At the early stages of this work we profited a lot from discussions with A.
Petkou.\\ G. A. is grateful to S. Frolov, S. Kuzenko and S. Theisen for useful
discussions. E. S. wishes to thank V. Dobrev, S. Ferrara, V. Petkova and I.
Todorov for enlightening discussions.\\ G. A. was supported by the DFG and by
the European Commission RTN programme HPRN-CT-2000-00131 in which he is
associated to U. Bonn, and in part by RFBI grant N99-01-00166 and by
INTAS-99-1782.

\renewcommand{\baselinestretch}{0.6}

\end{document}